\documentclass[aps,prd,onecolumn,groupedaddress]{revtex4}
\usepackage{graphicx}

\begin{document}

\title{Hubble Imaging Excludes Cosmic String Lens}
\author{Eric Agol, Craig J. Hogan, and Richard M. Plotkin}
\affiliation{Astronomy   Department, 
University of Washington,
Seattle, Washington 98195-1580}

\date{March 30, 2006}

\begin{abstract}
The galaxy image pair CSL-1 has been a leading  candidate for a cosmic string lens. High quality imaging data from the Hubble Space Telescope  presented here show that it is not a lens but a pair of galaxies. The galaxies show different orientations of their principal axes, not consistent with any lens model.
We present a new direct test of the  straight-string lens model, using
a displaced difference of the image from itself to exclude
CSL-1 at high confidence. 
\end{abstract}


\maketitle

Cosmic strings have been considered for 30 years as a possible new
form of cosmic mass-energy  \cite{kibble,zeldovich,vilenkin}.  Originally conceived as
vacuum defects in the context of field theories with broken
symmetries,  they are now sought as a possible macroscopic
manifestation of fundamental strings \cite{polchinski}.  In both cases,  they are often natural products of
inflationary cosmological models, and their predicted astrophysical behavior  has
been extensively studied  \cite{vilenkinshellard}.  Although
strings are now known not to play a dominant role in cosmic structure
formation, they may still create a rich observable phenomenology in
gravitational waves and
gravitational lensing. 

Over the years, several systems have been proposed as candidates for
string gravitational lenses. Most recently, the galaxy image pair
CSL-1 was proposed as a string lens \cite{sazhin1,sazhin2} generating
considerable excitement in the string community \cite{davis,kibble2,
  vilenkin2}. In this note, we present new data from the Hubble Space
Telescope (HST) 
awarded as Guest Observer time for Cycle 14 in April 2005 (GO-10486)
taken in February 2006.  We argue that CSL-1 is definitely not a
string lens, but likely a pair of galaxies, possibly in the early
stages of merging.  This
    conclusion agrees with independent results derived from Director's
    Discretionary time on HST awarded in May 2005 and taken in January
    2006 to address the same problem as our GO proposal \cite{Sazhin:2006fe}.

CSL-1 was discovered 
in  the  Capodimonte Deep Field  survey, where it appears as a pair of giant elliptical galaxy images separated by 1.9 arcsec. In ground based data, the two images show 
nearly identical morphologies, luminosities (M$_R\sim -22$), redshifts
($z=0.463 \pm 0.008,~\Delta v < 14 \pm 30$~km s$^{-1}$), and spectra \cite{sazhin1, sazhin2}. Although luminous ellipticals are often found in groups with close companions,  the interpretation presented, based on the very close resemblance,  was that  the two images are formed from lensing of a single galaxy by the conical defect in spacetime produced by a  nearly straight cosmic string lying between the images \cite[e.g.,][]{fairbairn}. 

We imaged CSL-1 with the {\it Wide-Field Channel} (WFC) on the
{\it Advanced Camera for Surveys} (ACS) for 5062 sec in F625W (sdds {\it r}) and
  2409 sec in F775W (sdss {\it i}), using a 4-point
    dither pattern in F625W and a 3-point dither pattern in F775W to
    reject hot pixels and cosmic rays.  The data was calibrated and
    exposures were combined using the standard HST reduction pipeline (e.g., {\it
      calacs} and {\it multidrizzle}).  This new data shows strong
    evidence against a cosmic string interpretation of CSL-1.  Direct examination of the image pair
    (Fig. 1) shows  that the two galaxies are indeed  similar
    elliptical galaxies, but their principal axes are significantly
    misaligned. This effect is not expected for  any kind of
    gravitational lens. Since the very similar redshifts indicate  the
    galaxies are in the same group, the  hint of asymmetry from  tidal
    streams forming in the outer parts of the images  suggests  that
    the two galaxies are in the early stages of a merger. 

 We
further test the string model using a nonparametric test.  A straight-string lens duplicates galaxy images in a strip of sky,
with no image distortion or parity reversal in a local region.  We
subtract the sky image  from a duplicate  of  itself, cut along the
best-fit projected string location and uniformly translated. The
projected location of the string, and the width of the duplicated
strip, introduce three parameters; if the data are well fit by a
cosmic string, the difference image should be consistent  with zero,
and the  reduced $\chi^2$ should be close to unity. Instead, we find
that the best-fit string-model difference image retains considerable
residual flux, as well as structure showing that  the images differ
from each other significantly  in their overall light distribution
(see Fig. 2).  We obtain a reduced $\chi^2$ of 5.8 and 5.0 in the
F625W and F775W difference images respectively, excluding the string model at a very high formal level of significance ($\approx 120\sigma$). 

%

\begin{acknowledgments}
Observations made with the NASA/ESA Hubble Space Telescope. Support for program \#10486 was provided by NASA  through a grant from  the Space Telescope Science Institute, which is operated by the Association of Universities for Research in Astronomy, Inc., under NASA contract NAS 5-26555.  This work was also supported by NSF grant AST-0098557 at the University of
Washington.
\end{acknowledgments}


 \begin{figure}
 \includegraphics[scale=0.75]{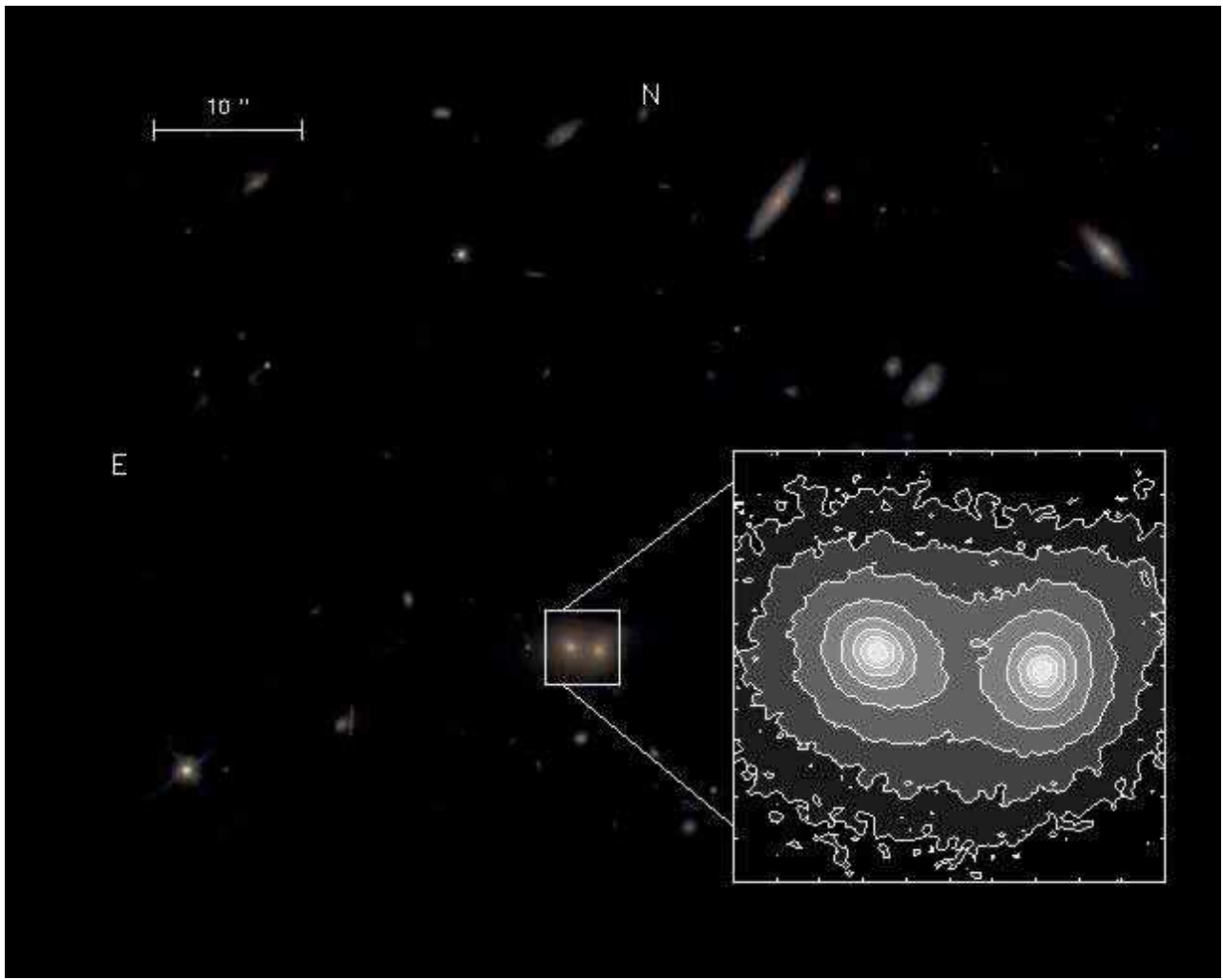}%
 \caption{Color composite of HST imaging data, generated using tools
  described by R. Lupton et al., Pub. Astron. Soc. Pac. 116, 133
  (2004). Inset shows a logarithmic contour
  plot of surface brightness in F775W for the CSL-1 galaxy pair;  contours  are separated by a factor  of 1.88
  in surface brightness, and the brightest corresponds to 19.405 mag
  per square arcsec.   The orientations of the isophotes are clearly
  different, strong evidence against a string.  The position angles of
  the  principal axes, measured   eastward from north,
  are  $51.7\pm 1.1$ deg for the eastern component and $-2.6\pm 1.5$
  deg for the  western component.}
 \end{figure}

\begin{figure}
\includegraphics[scale=0.75]{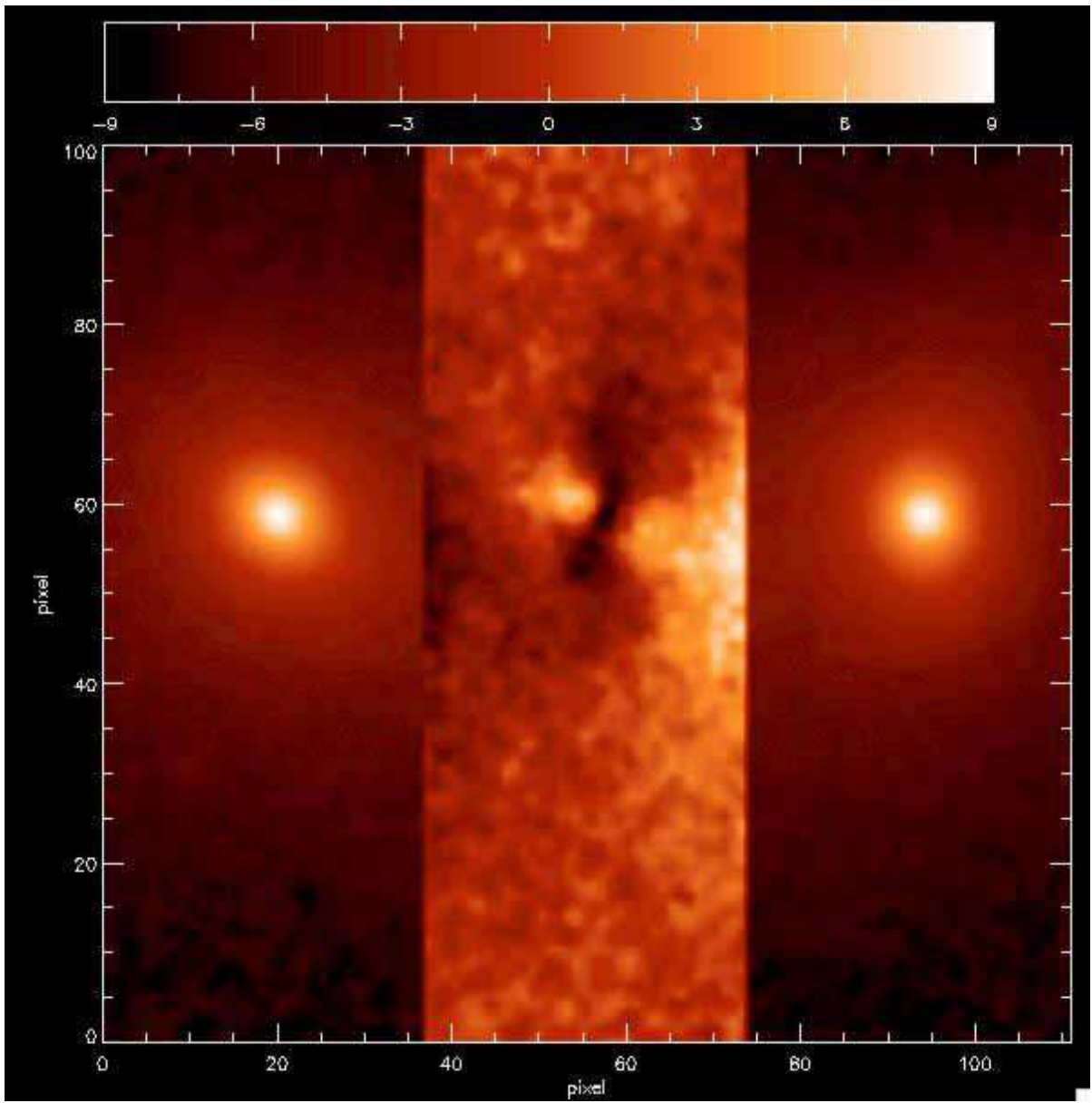}
\caption{Left and right panels show  HST F625W images of the
  two galaxies, with angular scale in units of $0.05$ arc sec
  pixels. The center panel shows the difference between the two for
  the best-fit cosmic string lens model. The figure
  shows the galaxies rotated so that the x-axis aligns with the
  translation axis (perpendicular to the ``string" angle).  The color bar refers to the
  center panel and indicates the flux difference in units of standard
  deviations per pixel.  The best-fit straight-string model gives a
  translation of 37.9 pixels at an angle of -7.1 degrees.  The reduced
  $\chi^2$ for the F625W difference image is 5.8 and for the F775W
  image is 5.0 within a strip 37 pixels wide by 100 pixels long. The
  string model predicts a null difference image, with reduced
  $\chi^2=1$ to high accuracy. }
\end{figure}

\end{document}